\documentclass[12pt,a4paper]{article}

\begin{document}

\title{\textbf{Coupled higher-order}\\%
\textbf{nonlinear Schr\"{o}dinger equations:}\\%
\textbf{a new integrable case via}\\%
\textbf{the singularity analysis}}
\author{\textsc{S. Yu. Sakovich}\\%
{\small Institute of Physics,}\\%
{\small National Academy of Sciences,}\\%
{\small P.O. 72, Minsk, BELARUS.}\\%
{\small sakovich@dragon.bas-net.by}
\and \textsc{Takayuki Tsuchida}\\%
{\small Department of Physics,\vspace{-0.1in}}\\%
{\small Graduate School of Science,\vspace{-0.1in}}\\%
{\small University of Tokyo,\vspace{-0.1in}}\\%
{\small Hongo 7-3-1, Bunkyo-ku,\vspace{-0.1in}}\\%
{\small Tokyo 113-0033, JAPAN.\vspace{-0.1in}}\\%
{\small tsuchida@monet.phys.s.u-tokyo.ac.jp}}
\date{\textsl{February 16, 2000}}
\maketitle
\begin{abstract}
A new system of coupled higher-order nonlinear Schr\"{o}dinger equations is
proposed which passes the Painlev\'{e} test for integrability well. A Lax pair
and a multi-field generalization are obtained for the new system.
\end{abstract}

\section{Introduction}

In this paper, we study the integrability of the following system of two
symmetrically coupled higher-order nonlinear Schr\"{o}dinger equations:%
\begin{equation}%
\begin{array}
[c]{l}%
q_{t}=hq_{xxx}+aq\bar{q}q_{x}+bq^{2}\bar{q}_{x}+cr\bar{r}q_{x}+dq\bar{r}%
r_{x}+eqr\bar{r}_{x}+\\
\multicolumn{1}{c}{\mathrm{i}(sq_{xx}+fq^{2}\bar{q}+gqr\bar{r}),\smallskip}\\
r_{t}=hr_{xxx}+ar\bar{r}r_{x}+br^{2}\bar{r}_{x}+cq\bar{q}r_{x}+dr\bar{q}%
q_{x}+erq\bar{q}_{x}+\\
\multicolumn{1}{c}{\mathrm{i}(sr_{xx}+fr^{2}\bar{r}+grq\bar{q}),}%
\end{array}
\label{sys}%
\end{equation}
where $h,a,b,c,d,e,s,f,g$ are real parameters, $h\neq0$, and the bar denotes
the complex conjugation.

Section \ref{singan} is devoted to the singularity analysis of (\ref{sys}),
and there, under certain simplifying assumptions, we find the following new
case when the system (\ref{sys}) passes the Painlev\'{e} test for
integrability well:%
\begin{equation}
a\neq0,\quad b=0,\quad c=d=e=a,\quad f=g=\frac{as}{3h}. \label{our}%
\end{equation}

In Section \ref{laxgen}, we prove the integrability of (\ref{sys}) with
(\ref{our}), constructing a corresponding $4\times4$ Lax pair, and then obtain
a multi-field generalization of the new integrable system.

Section \ref{discus} contains some concluding remarks.

\section{Singularity analysis\label{singan}}

Let us apply the Weiss-Kruskal algorithm of the singularity analysis
\cite{WTC}, \cite{JKM} to the system (\ref{sys}) (we set $h=1$ w.l.g.). With
respect to $q,\bar{q},r,\bar{r}$, which should be considered as mutually
independent during the Painlev\'{e} test, the system (\ref{sys}) is a normal
system of four third-order equations, of total order twelve. A hypersurface
$\phi(x,t)=0$ is non-characteristic for (\ref{sys}) if $\phi_{x}\neq0$, and we
set $\phi_{x}=1$. Then we substitute the expansions%
\begin{equation}%
\begin{array}
[c]{l}%
q=q_{0}(t)\phi^{\alpha}+\ldots+q_{n}(t)\phi^{n+\alpha}+\ldots,\\
\bar{q}=\bar{q}_{0}(t)\phi^{\beta}+\ldots+\bar{q}_{n}(t)\phi^{n+\beta}%
+\ldots,\\
r=r_{0}(t)\phi^{\gamma}+\ldots+r_{n}(t)\phi^{n+\gamma}+\ldots,\\
\bar{r}=\bar{r}_{0}(t)\phi^{\delta}+\ldots+\bar{r}_{n}(t)\phi^{n+\delta
}+\ldots
\end{array}
\label{exp}%
\end{equation}
(the bar does not mean the complex conjugation now) into (\ref{sys}) and
obtain four algebraic equations for $\alpha,\beta,\gamma,\delta,q_{0}\bar
{q}_{0},r_{0}\bar{r}_{0}$, which determine the dominant behavior of solutions
near $\phi=0$, as well as one twelfth-degree algebraic equation with respect
to $n$, which determines the positions of resonances in the expansions. The
perfect analysis of those five equations is very complicated, and we will
publish it later on. But now we impose certain restrictions on the expansions
(\ref{exp}) in order to reach a new integrable case of (\ref{sys}) by a short way.

For this purpose, we set $\alpha=\beta=\gamma=\delta=-1$ and require that
exactly two of twelve resonances lie in the position $n=0$. Under these
simplifying assumptions, we find from (\ref{sys}) and (\ref{exp}) that
$q_{0}\bar{q}_{0}=r_{0}\bar{r}_{0}=\mathrm{constant}\neq0$ (we set
$\mathrm{constant}=1$ w.l.g.), and that%
\begin{equation}
a=-6-b-c-d-e, \label{eqa}%
\end{equation}%
\begin{equation}%
\begin{array}
[c]{c}%
(n+1)n^{2}(n-3)(n-4)\times\\
(n^{2}-6n-2b-2d+5)(n^{2}-6n-2b-2e+5)\times\\
(n^{3}-6n^{2}+(5-2d-2e)n+4(c+d+e+3))=0.
\end{array}
\label{res}%
\end{equation}
Due to (\ref{res}), five resonances lie in the positions $n=-1,0,0,3,4$.
Denoting the positions of other seven resonances as $n_{1},n_{2},\ldots,n_{7}%
$, we find from (\ref{res}) that%
\begin{equation}%
\begin{array}
[c]{c}%
n_{2}=6-n_{1},\quad n_{4}=6-n_{3},\quad n_{7}=6-n_{5}-n_{6},\smallskip\\
d=\frac{1}{2}(5-2b-6n_{1}+n_{1}^{2}),\quad e=\frac{1}{2}(5-2b-6n_{3}+n_{3}%
^{2}),\smallskip\\
b=\frac{1}{4}(5-6n_{1}+n_{1}^{2}-6n_{3}+n_{3}^{2}+6n_{5}-n_{5}^{2}%
+6n_{6}-n_{6}^{2}-n_{5}n_{6}),\smallskip\\
c=\frac{1}{4}(-22+12n_{5}-2n_{5}^{2}+12n_{6}-2n_{6}^{2}-8n_{5}n_{6}+n_{5}%
^{2}n_{6}+n_{5}n_{6}^{2}).
\end{array}
\label{coe}%
\end{equation}
We require that the considered branch is generic, i.e. eleven resonances lie
in nonnegative positions. Taking into account the admissible multiplicity of
resonances, we have to study 23 distinct cases listed in Table \ref{tab}.%

\begin{table}[tbp] \centering
\begin{tabular}
[c]{||c||c|c|c|c||c||c||}\hline\hline
case \# & $n_{1}$ & $n_{3}$ & $n_{5}$ & $n_{6}$ & all resonances & $n_{log}%
$\\\hline\hline
1 & 1 & 1 & 1 & 1 & $-1,0,0,1,1,1,1,3,4,4,5,5$ & 1\\\hline
2 & 1 & 1 & 1 & 2 & $-1,0,0,1,1,1,2,3,3,4,5,5$ & 2\\\hline
3 & 1 & 1 & 2 & 2 & $-1,0,0,1,1,2,2,2,3,4,5,5$ & 2\\\hline
4 & 1 & 2 & 1 & 1 & $-1,0,0,1,1,1,2,3,4,4,4,5$ & 1\\\hline
5 & 1 & 2 & 1 & 2 & $-1,0,0,1,1,2,2,3,3,4,4,5$ & 2\\\hline
6 & 1 & 2 & 2 & 2 & $-1,0,0,1,2,2,2,2,3,4,4,5$ & 2\\\hline
7 & 1 & 3 & 1 & 1 & $-1,0,0,1,1,1,3,3,3,4,4,5$ & 1\\\hline
8 & 1 & 3 & 1 & 2 & $-1,0,0,1,1,2,3,3,3,3,4,5$ & 3\\\hline
9 & 1 & 3 & 2 & 2 & $-1,0,0,1,2,2,2,3,3,3,4,5$ & 2\\\hline
10 & 2 & 1 & 1 & 1 & $-1,0,0,1,1,1,2,3,4,4,4,5$ & 1\\\hline
11 & 2 & 1 & 1 & 2 & $-1,0,0,1,1,2,2,3,3,4,4,5$ & 2\\\hline
12 & 2 & 1 & 2 & 2 & $-1,0,0,1,2,2,2,2,3,4,4,5$ & 2\\\hline
13 & 2 & 2 & 1 & 1 & $-1,0,0,1,1,2,2,3,4,4,4,4$ & 1\\\hline
14 & 2 & 2 & 1 & 2 & $-1,0,0,1,2,2,2,3,3,4,4,4$ & NO, if (\ref{fgs})\\\hline
15 & 2 & 3 & 1 & 1 & $-1,0,0,1,1,2,3,3,3,4,4,4$ & 1\\\hline
16 & 2 & 3 & 1 & 2 & $-1,0,0,1,2,2,3,3,3,3,4,4$ & 3\\\hline
17 & 2 & 3 & 2 & 2 & $-1,0,0,2,2,2,2,3,3,3,4,4$ & 2\\\hline
18 & 3 & 1 & 1 & 1 & $-1,0,0,1,1,1,3,3,3,4,4,5$ & 1\\\hline
19 & 3 & 1 & 1 & 2 & $-1,0,0,1,1,2,3,3,3,3,4,5$ & 3\\\hline
20 & 3 & 1 & 2 & 2 & $-1,0,0,1,2,2,2,3,3,3,4,5$ & 2\\\hline
21 & 3 & 2 & 1 & 1 & $-1,0,0,1,1,2,3,3,3,4,4,4$ & 1\\\hline
22 & 3 & 2 & 1 & 2 & $-1,0,0,1,2,2,3,3,3,3,4,4$ & 3\\\hline
23 & 3 & 2 & 2 & 2 & $-1,0,0,2,2,2,2,3,3,3,4,4$ & 2\\\hline\hline
\end{tabular}
\caption{Positions of resonances and logarithmic terms.\label{tab}}%
\end{table}

At the next step of the analysis, we find from (\ref{sys}) and (\ref{exp}) the
recursion relations for $q_{n},\bar{q}_{n},r_{n},\bar{r}_{n}$, $n=0,1,2,\ldots
$, and then check the consistency of those relations at the resonances, using
the \emph{Mathematica} system \cite{Wol} for computations. The results are
listed in Table \ref{tab}, in its column $n_{log}$, where $n_{log}$ denotes
the position in which some logarithmic terms should be introduced into the
expansions (\ref{exp}) due to the following reasons: either the actual number
of arbitrary functions is less than the multiplicity of the resonance, or the
compatibility conditions at the resonance cannot be satisfied identically. As
we see, all the cases but one, \#14, have already failed to pass the
Painlev\'{e} test. In the case \#14, where%
\begin{equation}
h=1,\quad a=c=d=e=-\frac{3}{2},\quad b=0 \label{n14}%
\end{equation}
due to our assumption and formulae (\ref{eqa}) and (\ref{coe}), we have to set%
\begin{equation}
f=g=-\frac{1}{2}s \label{fgs}%
\end{equation}
for the compatibility conditions at $n=2,3$ to become identities.

The system (\ref{sys}) with (\ref{n14}) and (\ref{fgs}) admits many branches,
i.e. kinds of expansions (\ref{exp}). We have already studied the generic
branch. Other branches either are Taylor expansions governed by the
Cauchy-Kovalevskaya theorem, or are related to the following two:

\begin{enumerate}
\item $\alpha=-1$, $\beta=-1$, $\gamma=-2$, $\delta=2$, $q_{0}\bar{q}_{0}=4$,
$\forall\,r_{0},\bar{r}_{0}$, positions of resonances are
$n=-4,-1,0,0,0,1,1,3,4,4,5,5$;

\item $\alpha=-2$, $\beta=0$, $\gamma=-3$, $\delta=2$, $q_{0}\bar{q}_{0}=8$,
$\forall\,r_{0},\bar{r}_{0}$, positions of resonances are
$n=-5,-2,-1,0,0,0,2,4,5,5,6,7$.
\end{enumerate}

Compatibility conditions at all resonances of these branches turn out to be
identities. The Painlev\'{e} test is completed.

Since we consider $q$ and $\bar{q}$, $r$ and $\bar{r}$ as mutually
independent, evident scale transformations of $q,\bar{q},r,\bar{r}$ and $t$
relate the conditions (\ref{n14}) and (\ref{fgs}) with the more general
condition (\ref{our}). On the other hand, the transformation%
\begin{equation}%
\begin{array}
[c]{c}%
q^{\prime}=\frac{1}{2}q\exp\omega,\quad\bar{q}^{\prime}=-\frac{1}{2}\bar
{q}\exp(-\omega),\smallskip\\
r^{\prime}=\frac{1}{2}r\exp\omega,\quad\bar{r}^{\prime}=-\frac{1}{2}\bar
{r}\exp(-\omega),\smallskip\\
\omega=\frac{\mathrm{i}s}{3}x+\frac{2\mathrm{i}s^{3}}{27}t,\;x^{\prime
}=x+\frac{s^{2}}{3}t,\;t^{\prime}=-t
\end{array}
\label{tra}%
\end{equation}
changes the system (\ref{sys}) with (\ref{n14}) and (\ref{fgs}) into the
system of four coupled mKdV equations%
\begin{equation}%
\begin{array}
[c]{l}%
q_{t}+q_{xxx}+6q\bar{q}q_{x}+6(qr\bar{r})_{x}=0,\\
\bar{q}_{t}+\bar{q}_{xxx}+6q\bar{q}\bar{q}_{x}+6(\bar{q}r\bar{r})_{x}=0,\\
r_{t}+r_{xxx}+6r\bar{r}r_{x}+6(q\bar{q}r)_{x}=0,\\
\bar{r}_{t}+\bar{r}_{xxx}+6r\bar{r}\bar{r}_{x}+6(q\bar{q}\bar{r})_{x}=0,
\end{array}
\label{mkv}%
\end{equation}
where the prime of $x,t,q,\bar{q},r,\bar{r}$ is omitted, and the bar does not
mean (but may mean) the complex conjugation. This form (\ref{mkv}) is useful
for obtaining a Lax pair for the new integrable case (\ref{our}) of (\ref{sys}).

\section{Lax pair and generalization\label{laxgen}}

Let us consider the linear problem%
\begin{equation}
\Psi_{x}=U\Psi,\quad\Psi_{t}=V\Psi\label{lin}%
\end{equation}
with the matrices $U$ and $V$ given in the following block form \cite{TW}:%
\begin{equation}
U=\mathrm{i}\zeta\left(
\begin{array}
[c]{cc}%
-I_{1} & 0\\
0 & I_{2}%
\end{array}
\right)  +\left(
\begin{array}
[c]{cc}%
0 & Q\\
R & 0
\end{array}
\right)  , \label{bfu}%
\end{equation}%
\begin{equation}%
\begin{array}
[c]{c}%
V=\mathrm{i}\zeta^{3}\left(
\begin{array}
[c]{cc}%
-4I_{1} & 0\\
0 & 4I_{2}%
\end{array}
\right)  +\zeta^{2}\left(
\begin{array}
[c]{cc}%
0 & 4Q\\
4R & 0
\end{array}
\right)  +\\
\mathrm{i}\zeta\left(
\begin{array}
[c]{cc}%
-2QR & 2Q_{x}\\
-2R_{x} & 2RQ
\end{array}
\right)  +\left(
\begin{array}
[c]{cc}%
Q_{x}R-QR_{x} & -Q_{xx}+2QRQ\\
-R_{xx}+2RQR & R_{x}Q-RQ_{x}%
\end{array}
\right)  ,
\end{array}
\label{bfv}%
\end{equation}
where $I_{1}$ and $I_{2}$ are unit matrices, $\zeta$ is a parameter. The
compatibility condition of the linear problem (\ref{lin}),%
\begin{equation}
U_{t}-V_{x}+UV-VU=0, \label{com}%
\end{equation}
becomes the system of two matrix mKdV equations \cite{AF}:%
\begin{equation}%
\begin{array}
[c]{l}%
Q_{t}+Q_{xxx}-3Q_{x}RQ-3QRQ_{x}=0,\\
R_{t}+R_{xxx}-3R_{x}QR-3RQR_{x}=0.
\end{array}
\label{meq}%
\end{equation}
If we substitute%
\begin{equation}
Q=\left(
\begin{array}
[c]{cc}%
q & r\\
\bar{r} & \bar{q}%
\end{array}
\right)  ,\quad R=-\left(
\begin{array}
[c]{cc}%
\bar{q} & r\\
\bar{r} & q
\end{array}
\right)  \label{sqr}%
\end{equation}
into (\ref{meq}), we obtain exactly the new system (\ref{mkv}). This proves
that the new case (\ref{our}) of the coupled higher-order nonlinear
Schr\"{o}dinger equations (\ref{sys}) possesses a parametric Lax pair.

A multi-field generalization of the system (\ref{mkv}) is obtained by choosing%
\begin{equation}%
\begin{array}
[c]{l}%
Q=\left(
\begin{array}
[c]{cc}%
u_{0}I\otimes I+\sum_{k=1}^{2m-1}u_{k}e_{k}\otimes I & v_{0}I\otimes
I+\sum_{k=1}^{2m-1}v_{k}I\otimes e_{k}\\
v_{0}I\otimes I-\sum_{k=1}^{2m-1}v_{k}I\otimes e_{k} & u_{0}I\otimes
I-\sum_{k=1}^{2m-1}u_{k}e_{k}\otimes I
\end{array}
\right)  ,\smallskip\\
R=-\left(
\begin{array}
[c]{cc}%
u_{0}I\otimes I-\sum_{k=1}^{2m-1}u_{k}e_{k}\otimes I & v_{0}I\otimes
I+\sum_{k=1}^{2m-1}v_{k}I\otimes e_{k}\\
v_{0}I\otimes I-\sum_{k=1}^{2m-1}v_{k}I\otimes e_{k} & u_{0}I\otimes
I+\sum_{k=1}^{2m-1}u_{k}e_{k}\otimes I
\end{array}
\right)  ,
\end{array}
\label{gqr}%
\end{equation}
where $I$ is the $2^{m-1}\times2^{m-1}$ unit matrix, and $\{e_{1}%
,\ldots,e_{2m-1}\}$ are $2^{m-1}\times2^{m-1}$ anti-commutative and
anti-Hermitian matrices:%
\begin{equation}
\{e_{i},e_{j}\}_{+}=-2\delta_{ij}I,\quad e_{k}^{\dagger}=-e_{k}. \label{eij}%
\end{equation}
Then the compatibility condition (\ref{com}) becomes the system%
\begin{equation}%
\begin{array}
[c]{l}%
u_{j,t}+u_{j,xxx}+6\sum_{k=0}^{2m-1}u_{k}^{2}u_{j,x}+6\left(  \sum
_{k=0}^{2m-1}v_{k}^{2}u_{j}\right)  _{x}=0,\smallskip\\
v_{j,t}+v_{j,xxx}+6\sum_{k=0}^{2m-1}v_{k}^{2}v_{j,x}+6\left(  \sum
_{k=0}^{2m-1}u_{k}^{2}v_{j}\right)  _{x}=0,\smallskip\\
\multicolumn{1}{r}{j=0,1,\ldots,2m-1.}%
\end{array}
\label{muv}%
\end{equation}
If we assume that $u_{k}$ and $v_{k}$ are real and set%
\begin{equation}%
\begin{array}
[c]{rc}%
\begin{array}
[c]{c}%
u_{2j-2}+\mathrm{i}u_{2j-1}=q_{j},\\
v_{2j-2}+\mathrm{i}v_{2j-1}=r_{j},
\end{array}
& j=1,2,\ldots,m,
\end{array}
\label{cfy}%
\end{equation}
the system (\ref{muv}) is expressed as%
\begin{equation}%
\begin{array}
[c]{l}%
q_{j,t}+q_{j,xxx}+6\sum_{k=1}^{m}\left|  q_{k}\right|  ^{2}q_{j,x}+6\left(
\sum_{k=1}^{m}\left|  r_{k}\right|  ^{2}q_{j}\right)  _{x}=0,\smallskip\\
r_{j,t}+r_{j,xxx}+6\sum_{k=1}^{m}\left|  r_{k}\right|  ^{2}r_{j,x}+6\left(
\sum_{k=1}^{m}\left|  q_{k}\right|  ^{2}r_{j}\right)  _{x}=0,\smallskip\\
\multicolumn{1}{r}{j=1,2,\ldots,m.}%
\end{array}
\label{mqr}%
\end{equation}

\section{Conclusion\label{discus}}

In the literature, the following three integrable cases of coupled
higher-order nonlinear Schr\"{o}dinger equations (\ref{sys}) are known and
studied (sometimes in a form of coupled mKdV equations):%
\begin{equation}
a\neq0,\quad b=e=0,\quad c=d=\frac{1}{2}a,\quad f=g=\frac{as}{3h},\label{ch1}%
\end{equation}%
\begin{equation}
a\neq0,\quad b=d=e=0,\quad c=a,\quad f=g=\frac{as}{3h},\label{ch2}%
\end{equation}%
\begin{equation}
a\neq0,\quad b=d=e=\frac{1}{3}a,\quad c=\frac{2}{3}a,\quad f=g=\frac{2as}%
{9h};\label{css}%
\end{equation}
they were introduced in \cite{TP}, \cite{IH}, \cite{PSM}, respectively.

The new integrable case (\ref{our}) of the system (\ref{sys}), obtained in
this paper by means of the singularity analysis, admits the reduction
$r\rightarrow0$ to the Hirota equation \cite{Hir} and the reduction
$r\rightarrow q$ to the Sasa-Satsuma equation \cite{SS}. Its soliton solutions
and conservation laws deserve further investigation.\bigskip

\textbf{Acknowledgments.} The work of S.~Yu.~S. was supported in part by the
Fundamental Research Fund of Belarus, grant $\Phi$98-044. The work of T.~T.
was supported by a JSPS Research Fellowship for Young Scientists.

\end{document}